\documentclass[runningheads, envcountsame, a4paper]{llncs}
\usepackage[pdftex]{graphicx}
\usepackage{epstopdf}
\usepackage{amsmath}
\usepackage{xcolor}
\usepackage{diagbox}
\usepackage{slashbox}
\usepackage{multirow}
\usepackage{float}
\usepackage[hyphens]{url}
\usepackage[misc]{ifsym}

\begin{document}
\title{Pushing the Limits of Exoplanet Discovery via Direct Imaging with Deep Learning}
\titlerunning{Direct Imaging of Exoplanets with Deep Learning}
%
\author{Kai Hou Yip\inst{1} \Letter{}
\and
Nikolaos Nikolaou\inst{1} 
\and
Piero Coronica\inst{4}
\and
Angelos Tsiaras\inst{1}
\and
\\Billy Edwards \inst{1}
\and
Quentin Changeat \inst{1}
\and
Mario Morvan \inst{1}
\and
Beth Biller\inst{2}
\and
\\Sasha Hinkley\inst{3}
\and
Jeffrey Salmond\inst{4}
\and
Matthew Archer\inst{4}
\and
Paul Sumption\inst{4}
\and 
\\Elodie Choquet\inst{5}
\and
Remi Soummer\inst{6}
\and
Laurent Pueyo\inst{6}
\and
Ingo P. Waldmann\inst{1}}
\authorrunning{K. H. Yip et al.}
\tocauthor{K. H. Yip et al.}
\toctitle{Pushing the Limits of Exoplanet Discovery via Direct Imaging with Deep Learning}
%
\institute{Department of Physics and Astronomy, University College London,\\Gower Street, London, WC1E 6BT, UK \\
\email{$\{$kai.yip.13, n.nikolaou, angelos.tsiaras.14, billy.edwards.16, quentin.changeat.18, mario.morvan.18, ingo.star$\}$@ucl.ac.uk}\\
\and
SUPA, Institute for Astronomy; Centre for Exoplanet Science,\\University of Edinburgh, Edinburgh, UK\\
\email{bb@roe.ac.uk}\\
\and
Department of Physics and Astronomy, University of Exeter,\\Stocker Road, EX4 4PY, UK\\
\email{s.hinkley@exeter.ac.uk}
\and
Research Software Engineering, University of Cambridge,\\Trinity Lane, Cambridge CB2 1TN, UK \\
\email{$\{$pc620,js947,ma595,ps459$\}$@cam.ac.uk}\\
\and
Aix Marseille Univ, CNRS, CNES, LAM, Marseille, France \\
\email{elodie.choquet@lam.fr}\\
\and
STScI, 3700 San Martin Drive, Baltimore, MD 21218, U.S. \\
\email{$\{$soummer, pueyo$\}$@stsci.edu}
}
\maketitle 
\begin{abstract}
Further advances in exoplanet detection and characterisation require sampling a diverse population of extrasolar planets. 
One technique to detect these distant worlds is through the direct detection of their thermal emission. The so-called direct imaging technique, is suitable for observing young planets far from their star. 
These are very low signal-to-noise-ratio (SNR) measurements and limited ground truth hinders the use of supervised learning approaches. 
In this paper, we combine deep generative and discriminative models to bypass the issues arising when directly training on real data. We use a Generative Adversarial Network to obtain a suitable dataset for training Convolutional Neural Network classifiers to detect and locate planets across a wide range of SNRs. Tested on artificial data, our detectors exhibit good predictive performance and robustness across SNRs. To demonstrate the limits of the detectors, we provide maps of the precision and recall of the model per pixel of the input image. On real data, the models can re-confirm bright source detections. \\

\end{abstract}

\section{Introduction}

In the last 20 years, our understanding of planetary science has undergone what can best be described as a \emph{second Copernican revolution}. With over 4000 discovered \footnote{Paris Observatory Exoplanet Catalogue: \url{http://exoplanet.eu}} exoplanets  -- planets orbiting stars other than our sun -- we now understand that planet formation is an integral part of stellar formation. In other words, every star in our galaxy is likely to host at least one planet \cite{Cassan_pl_stat}. To date, we have only just begun to understand the mechanisms underlying planet formation, evolution and potential habitability and it is only by studying a large population of extrasolar planets that we can begin to place our own solar system in the galactic context. 
Hence, it is no surprise that the field of extrasolar planets is one of the fastest growing and most dynamic in contemporary astrophysics.


Several exoplanet detection techniques exist. Through their various observational constraints, we find each technique to be sensitive to a specific subset of the planet population. 
While most detection techniques are indirect in nature and only measure the planet's effect on the received stellar light, we here concern ourselves with the most direct detection method: \emph{direct imaging}. As the name suggests, direct imaging tries to image the planet's thermal emission in-situ by blocking out the light of its host-star to reveal the significantly fainter planetary companion. These directly imaged planets are very young as they still radiate from the heat of their recent formation \cite{2016direct_img_review,lagrange_betapic,kalas}.  Hence, studying this population gives us a window into early planet formation.
To understand the formation and evolution history of our own solar system, it is paramount to study the widest possible range of planetary systems and ages. Therefore, more detections via direct imaging would greatly impact the field. 


\subsection{The challenge of direct imaging}

Detecting a planet via direct imaging poses a significant challenge. Despite significant efforts and technological advances made with this technique, the number of confirmed detections -- only 16 exoplanets so far~\cite{2016direct_img_review} -- remains far behind those of other methods. The object of interest is often significantly dimmer than the parent star (best case contrast ratio of $\sim10^{-5}$). In practice, astronomers increase the planet-flux\footnote{The term \emph{flux} refers to the rate of incoming photons.} contrast using a \emph{coronagraph}, a mask blocking the star's light. The resulting image has a reduced contribution from the stellar flux but is still subject to systematic residual flux by the diffracted stellar light inside the instrument optics, resulting in a low \emph{Signal-to-Noise Ratio (SNR)} per image. This systematic noise pattern is known as the `\emph{speckle noise}' and is an instrument specific, quasi-static pattern of light on the detector \cite{racine1999speckle}.

To first order the quasi-static nature of the speckle noise allows it to be removed by \emph{image subtraction}. This is often performed using \emph{Angular Differential Imaging (ADI)}, where a sequence of images are taken at different roll-angles (i.e. orientations) of the telescope. However for space-based observatories, such as the \emph{Hubble Space Telescope (HST)} or the upcoming \emph{James Web Space Telescope (JWST)}, choices of roll-angles are limited and calibration images are expensive to obtain. Hence we must use other speckle pattern suppression techniques.


The \emph{limited ground truth} (knowledge on whether a given system contains planets or not) has largely restricted any machine learning approaches applied to the problem to \emph{unsupervised learning} techniques.
By obtaining a low rank approximation of the data, one can compute the difference of an image and its reconstruction after being projected to this lower dimensional space to remove the dominant components of the speckle pattern. Such methods include, \emph{LOCI}~\cite{Loci}, \emph{principal component analysis (PCA)-based algorithms}~\cite{Soummer,Amara,choquet2018hd}, and \emph{LLSG}~\cite{LLSG} and can be used as a denoising step before \emph{visual inspection}. They cannot classify images as possibly containing planets or not, nor automatically locate planets in images. The latter is achievable by \emph{ANDROMEDA}~\cite{mugnier,cantalloube2015direct} via maximum likelihood estimation on the residual images obtained by pairwise subtraction within the ADI sequence but naturally, it is only applicable to ADI sequence data and hence not applicable to space-based observatories.

The number of \emph{supervised learning} attempts is limited. In \cite{fergus2014s4}, the authors use a \emph{Support Vector Machine (SVM)} to classify images as possibly containing planets or not. Here, the authors injected fake planets into real data without establishing the ``planet-free" ground-truth first. In \cite{gonzalez2018supervised}, the authors used \emph{Singular Value Decomposition (SVD)} of ADI sequence data to remove possible planet signals before planet injection. A \emph{Random Forest} and a \emph{Convolutional LSTM} classifier were then trained on the pre-processed data. Unfortunately, according to the authors, the method did not work on single images but only on sequences of ADI images. For space-based data considered in this paper, this method is therefore not applicable as often only individual images are obtained with space-based telescopes. Moreover, the results reported appear to suggest that their final models produce a very large number of false positive planet detections per frame. 




All these methods suffer from class imbalance, i.e. a lack of confirmed planet detections (positive examples) to train on.
However, in order to train a supervised model for exoplanet detection, we require a sufficient amount of examples from both classes. Similarly, images without currently known planets should not be used as negative examples. This is because it is possible --in fact, probable-- that undiscovered planets are present in the data. For these reasons it is imprudent to train supervised models directly on real data. Simply put, if one assumes `absence of detection' as `confirmation of absence' and uses this to train a new model, then (i) the performance of the model is upper bounded by the current state-of-the-art and (ii) the training dataset is biased towards the negative class.

\subsection{Overview of the paper}

In this work, we circumvent the aforementioned issues by introducing an intermediate step of generative modelling between the real data and the final discriminative model. By training a \emph{Generative Adversarial Network (GAN)} \cite{GAN} on real data from the \emph{Near Infrared Camera and Multi-Object Spectrometer (NICMOS)} on the HST, we obtain a generative model of the distribution of the most prominent component of the data: the \emph{speckle noise pattern}, which is the main component of systematic noise in direct imaging. At this stage, any planet signals in the training data are regarded as statistical noise since their occurrence is random and usually buried within the speckle pattern. Our generative model can produce negative class examples (images without planets). We can then create an equal amount of synthetic positive examples (images with planets) by `injecting' planets on images generated by the GAN. 

\begin{figure}
    \centering
    \includegraphics[width = 1\linewidth]{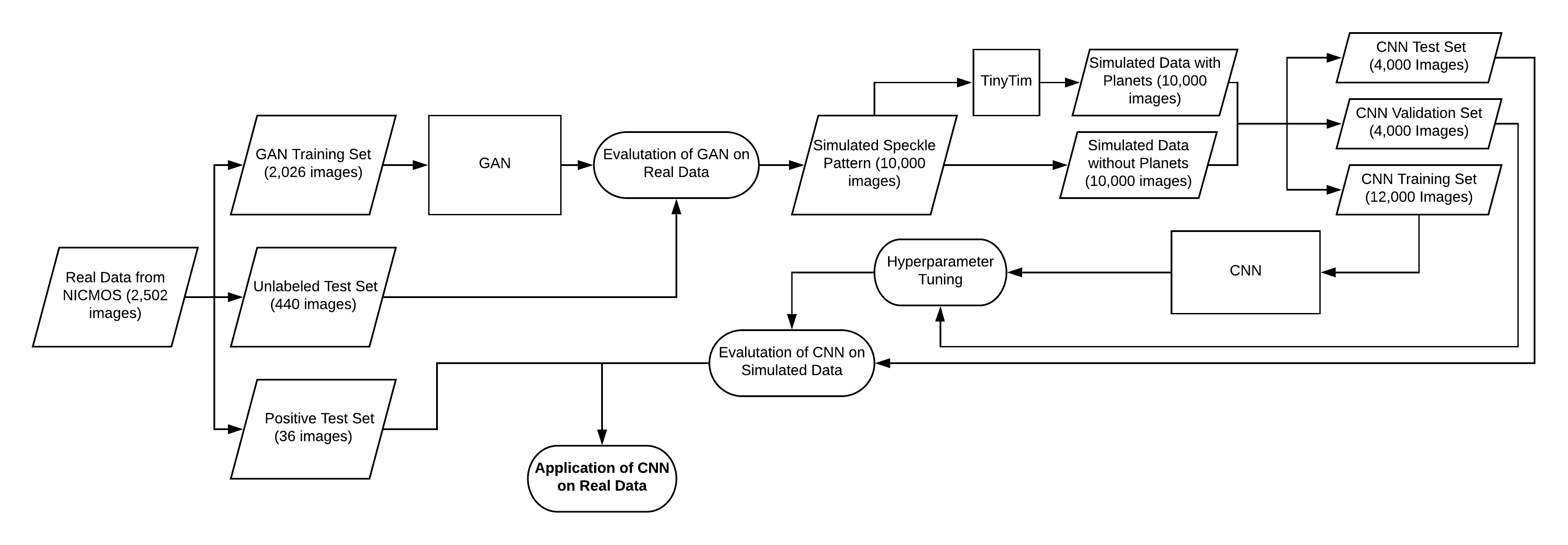}
    \caption{Flowchart summarising the methodology utilised in this paper.}
    \label{layout}
\end{figure}

We use this dataset to train a \emph{Convolutional Neural Network (CNN)} to classify images as positive/negative. By doing so, we avoid the problem of using the real data directly and all the issues that come with it (class imbalance, unknown ground truth, model's performance being upper-bounded by current detection techniques). With the use of \emph{Class Activation Maps (CAM)} \cite{zhou2016learning}, we are able to locate injected planets within the images, as a byproduct. Finally, we turn to real data for evaluation and demonstrate that our model can identify \emph{confirmed bright sources}\footnote{There are no confirmed planet detections on NICMOS filter F110W yet. These \emph{bright sources} are almost certainly background stars. However, detecting these showcases the potential for any bright source --including planets-- to be detected.} in the dataset. The architecture of our model is shown in Fig. \ref{layout}. 

\section{Data \& preprocessing}

\subsection{The dataset}

The dataset we used consists of single coronagraphic images taken by \emph{Camera 2} on the \emph{HST/NICMOS} instrument during \emph{Cycle 11-16 (Data Delivery 2)}\footnote{The original dataset is publicly available at the \emph{HST LAPLACE STScI} archive \url{https://archive.stsci.edu/prepds/laplace/}}. All images are from the same wavelength channel, \emph{filter F110W}, and have a dimension of $256\times256$ pixels\footnote{The \emph{Field of View (FOV)} of the camera, is 19.2"$\times$19.2" corresponding to images of size 256$\times$256 pixels and the \emph{coronagraph} is a circular disk with a radius of 4 pixels.}. The images have already undergone some standard image processing commonly applied to HST data (i.e. \emph{bias calibration}, \emph{dark calibration} \& \emph{flat calibration})\footnote{\emph{Bias calibration} removes unwanted saturated pixels that arise during long exposures. \emph{Dark calibration} corrects for thermal emissions coming from the detector. \emph{Flat calibration} corrects for differences in sensitivity across the CCD detector.}.
\emph{All} images have undergone the \emph{contemp\_flats} calibration, where a flat field lamp is used to create a reference image during the process of target acquisition. 

This means that the final dataset includes several images of the same observation (raw data) that have undergone different transformations. These transformations can be viewed as a form of \emph{data augmentation} of the raw data. To avoid overfitting, we were careful to include all images of the same target in either only the training set or only the test set. The augmented dataset used consists of 3572 single-channel (F110W filter) images.

\subsection{Preprocessing}

\begin{figure}
    \centering
    \includegraphics[width=0.45\linewidth]{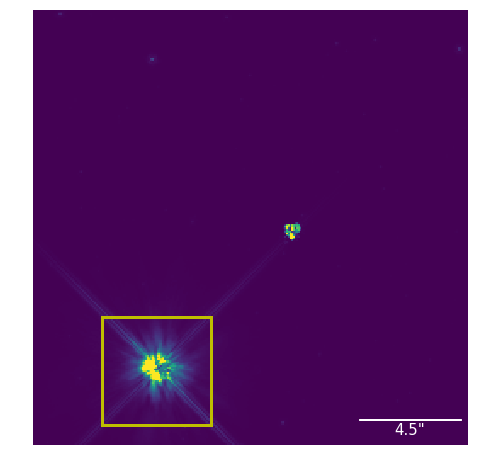}
    \centering
    \includegraphics[width=0.45\linewidth]{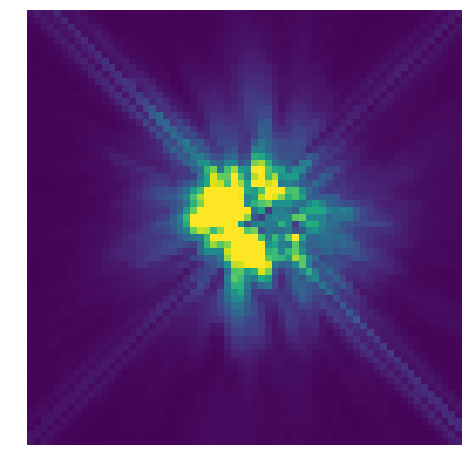}
    \caption{Left: Example of a full-sized 256$\times$256 image taken by NICMOS. Right: The cropped 64$\times$64 image (represented by the yellow square on the left) containing only the vicinity of the stellar speckle pattern.}
    \label{fig:crop_img}
\end{figure}

\emph{Image cropping:} The areas of the $256\times256$ images that interest us the most (i.e. those with the highest probability of detecting a planet) are regions in the vicinity of the star, and hence the \emph{speckle pattern}. This occupies only a small portion of the original image so each image was arbitarily cropped into a $64\times64$ frame. The cropping procedure is carried out by setting [40,180] as the top-left corner of the $64\times64$ box. As all images are aligned, the position of the speckle pattern stays the same. In Fig. \ref{fig:crop_img}, we give an example of an image before and after cropping.

\emph{Treatment of corrupted images:} Out of the initial 3572 images, some were found to contain \emph{overexposed pixels}, \emph{unexplained bright spots}\footnote{Large bright spots found outside the speckle pattern } or \emph{de-focused frames}. 
After removing these unsuitable frames from the dataset, we are left with 2502 images.

\emph{Normalization:} The images were normalised linearly so that their pixel intensities are in the range [0,1] using the maximum and minimum value of each pixel across the entire dataset. 

\emph{Final dataset \& train/test split:} The final dataset of real observations included 2502 single-channel $64\times64$ images with pixel intensities in [0,1]. 

We reserved 476 images from this dataset -- about 20\% of the available datapoints -- for evaluation purposes and use the remaining 2026 for training. The training set and the test set contain no targets in common. The test set contains 36 images of stars with confirmed bright sources (3 targets). The final dataset used in the paper is thus split into 3 parts: \emph{Training} (2026 real images w/o confirmed bright sources), \emph{Test (Positive)} (36 real images with confirmed bright sources) \& \emph{Test (Unlabeled)} (440 real images w/o confirmed bright sources)\footnote{Data and code are available at  \url{https://github.com/ucl-exoplanets/DI-Project}}.

\section{A GAN for modelling the speckle pattern distribution}

To train a discriminative model for identifying planets in direct imaging observations we require sufficient data from both classes. Yet positive examples, images containing confirmed exoplanets, are scarce (only 3 targets in our dataset contain confirmed bright sources, amounting to a total of 36 images). Even if there were more, a model trained to only identify planets \emph{detectable} by the current technology would not advance our existing detection capabilities. On the other hand, images without currently known planets cannot be used as negative examples, as they might contain \emph{undiscovered} planets. In fact, we \emph{expect} many of them to do so and the purpose of this paper is to further our ability to discover planets in such images; these examples are \emph{unlabelled}, not negative. Directly using data from real observations to train a discriminative model is therefore unjustified. 

Note however, that although we are unaware of the presence of undetected exoplanets in the original data, we know of one component of these images that is present in \emph{all} of them and hinders our ability to detect planets: the \emph{instrument speckle pattern}. Therefore we can instead use the original data to train a generative model of this pattern. This will be the first step towards generating an artificial labelled dataset. Negative examples will consist of instances of this pattern alone and positive examples will consist of the speckle pattern with the introduction of planet signals in the images. 


We train a \emph{Deep Convolutional Generative Adversarial Network (DCGAN)}~\cite{DCGAN} with a latent space of 100 dimensions to learn a model of the speckle pattern. We have chosen DCGAN as the base due to its stability during training, slightly modifying the architecture to alleviate the \emph{checkerboard effect}\footnote{The term refers to artifacts caused by the uneven overlap of the deconvolutions of a CNN when the kernel (filter) size is not divisible by the number of strides.}.

A detailed description of the architecture of the GAN can be found in Table~\ref{tab:GAN_config}. The GAN was trained for 40 epochs using the \emph{ADAM} optimiser with a batch size of 16 and a learning rate of $2\times10^{-4}$. The remaining hyperparameters were set to default \emph{Keras}\footnote{\url{https://keras.io}} values. The hyperparameter optimisation was based on minimising the validation loss on 20\% of the training set and was minimal due to the already good performance of the final model. A principled hyperparameter exploration is left for future work.

\begin{table}[t]
\vspace{-0.5cm}
\scriptsize
\centering
\begin{tabular}{ccc}
\multicolumn{3}{c}{Generator}                                                                                       \\ \hline 
\multicolumn{1}{c|}{Layer Type}        & \multicolumn{1}{c|}{Dimensionality, Configuration} & Output Dimension \\ \hline \hline
\multicolumn{1}{c|}{Latent Space}                 & \multicolumn{1}{c|}{}                                   & (m,100)          \\
\multicolumn{1}{c|}{FC-BN-RELU} & \multicolumn{1}{c|}{2048} & (m,4,4,128)      \\
\multicolumn{1}{c|}{Resize}            & \multicolumn{1}{c|}{}                                   & (m,8,8,128)      \\
\multicolumn{1}{c|}{Conv-BN-RELU}  & \multicolumn{1}{c|}{5$\times$5, f=64, s=1}  & (m,8,8,64)       \\
\multicolumn{1}{c|}{Resize}            & \multicolumn{1}{c|}{}                                   & (m,16,16,64)     \\
\multicolumn{1}{c|}{Conv-BN-RELU}  & \multicolumn{1}{c|}{5$\times$5, f=32, s=1}  & (m,16,16,32)     \\
\multicolumn{1}{c|}{Resize}            & \multicolumn{1}{c|}{}                                   & (m,32,32,32)     \\
\multicolumn{1}{c|}{Conv-BN-RELU}  & \multicolumn{1}{c|}{5$\times$5, f=16, s=1}  & (m,32,32,16)     \\
\multicolumn{1}{c|}{Resize}            & \multicolumn{1}{c|}{}                                   & (m,64,64,16)     \\
\multicolumn{1}{c|}{Conv-BN-RELU}  & \multicolumn{1}{c|}{5$\times$5, f=1, s=1}   & (m,64,64,1)      \\
\multicolumn{1}{c|}{Sigmoid}           & \multicolumn{1}{c|}{}                                   & (m,64,64,1)     \\

\multicolumn{3}{c}{Discriminator}                                                              \\ \hline \hline
\multicolumn{1}{c|}{Input}             & \multicolumn{1}{c|}{64$\times$64$\times$1}       & (64,64)             \\
\multicolumn{1}{c|}{Conv-LeakyRELU}   & \multicolumn{1}{c|}{3$\times$3, f=16, s=1}    & (m,64,64,16)     \\
\multicolumn{1}{c|}{MaxPool}           & \multicolumn{1}{c|}{2$\times$2, s=2}       & (m,32,32,16)     \\
\multicolumn{1}{c|}{Conv-BN-LeakyRELU} & \multicolumn{1}{c|}{3$\times$3, f=32, s=1}    & (m,32,32,32)     \\
\multicolumn{1}{c|}{MaxPool}           & \multicolumn{1}{c|}{2$\times$2, s=2}       & (m,16,16,32)     \\
\multicolumn{1}{c|}{Conv-BN-LeakyRELU} & \multicolumn{1}{c|}{3$\times$3, f=64, s=1}    & (m,16,16,64)     \\
\multicolumn{1}{c|}{MaxPool}           & \multicolumn{1}{c|}{2$\times$2, s=2}       & (m,8,8,64)       \\
\multicolumn{1}{c|}{Conv-BN-LeakyRELU} & \multicolumn{1}{c|}{3$\times$3, f=128, s=1}   & (m,8,8,128)      \\
\multicolumn{1}{c|}{FC}                & \multicolumn{1}{c|}{256}           & (m,256)          \\
\multicolumn{1}{c|}{Sigmoid}           & \multicolumn{1}{c|}{2}             & (m,2)  \\ \hline
\end{tabular}
\small
\caption{Architecture of the GAN used for generating the synthetic data. `BN', `Conv' \& `FC' denote Batch Normalisation, Convolutional \&  Fully Connected layers. We denote the \# of convolutional filters with `f', the stride size with `s' \& the batch size with `m'.}
\label{tab:GAN_config}
\vspace{-1cm}
\end{table}

After training the GAN by minimising the classical cross-entropy-based \emph{adversarial loss} \cite{GAN} on the training set, we evaluate the performance of the \emph{generator} on the unlabelled test data (440 images w/o confirmed bright sources), based on its ability to reconstruct them. For each of the test examples (real datapoints), we generate 600 samples by the generator (artificial datapoints) and following the approach of \cite{Yeh}, we select the one minimising $L_c +\lambda L_p$, where $L_c$ is the \emph{contextual loss} (measuring the difference between observed and generated data using the \emph{pixelwise L1-distance} between the two images), $L_p$ the \emph{perceptual loss} (uses the discriminator to verify the validity of the generated data given the training) and $\lambda$, here set to $0.1$, controls their trade-off. Comparing the two images gives us an indication of how well the GAN models the real data distribution.

\begin{figure}
    \vspace{-0.5cm}
    \centering
    \includegraphics[width=0.3\linewidth]{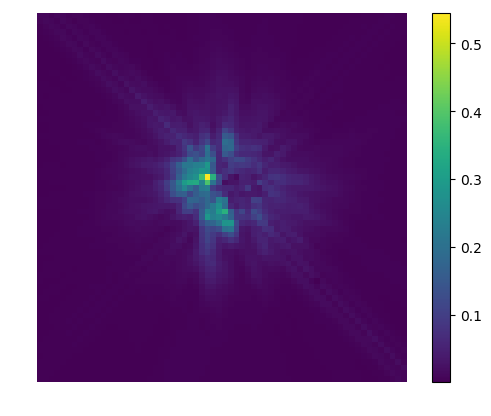}
    \includegraphics[width=0.3\linewidth]{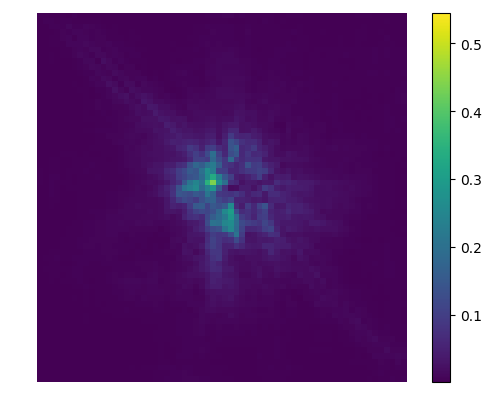}
    \includegraphics[width=0.3\linewidth]{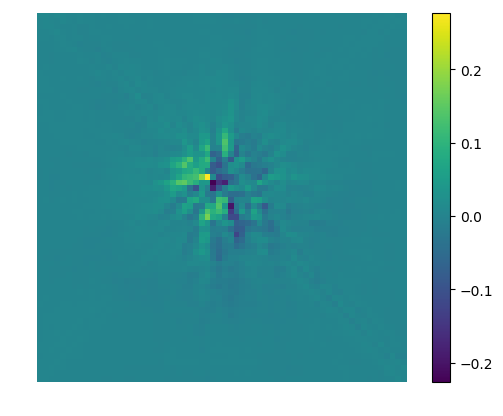}
    \includegraphics[width=0.3\linewidth]{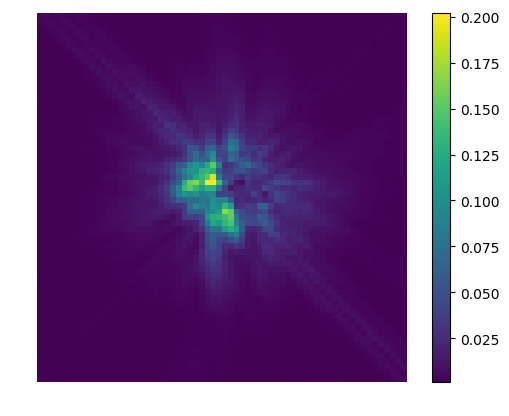}
    \includegraphics[width=0.3\linewidth]{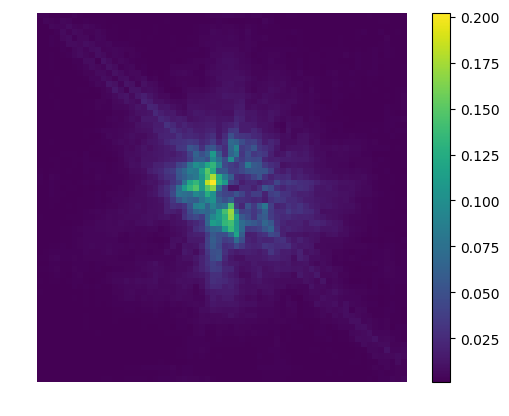}
    \includegraphics[width=0.3\linewidth]{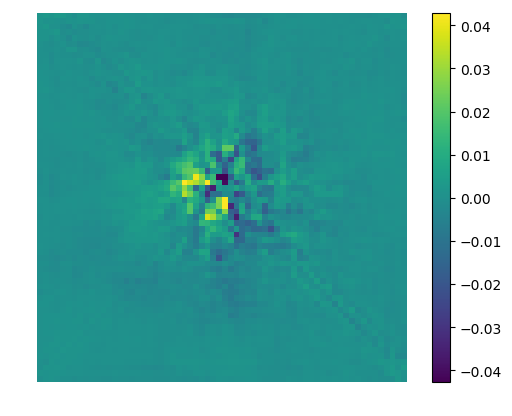}
    \vspace{-10pt}
    \caption{An original image [left], a reconstruction generated by the GAN [middle] and their difference [right]. Top: Worst reconstruction case in test set (dissimilarity score $6.7\times 10^{-3}$); Bottom: A reconstruction of average quality (dissimilarity score $2.0\times 10^{-3}$).}
    \label{fig:gan_compare}
\end{figure}

We can assess the quality of the reconstructed images qualitatively, by visual inspection of generated and original images, or quantitatively, by computing their dissimilarity (\emph{pixelwise L1-distance}). Fig. \ref{fig:gan_compare} shows examples of synthetic images generated by the GAN compared to their real image counterparts and their differences. We show the worst reconstruction produced by the GAN with a dissimilarity score of $6.7\times 10^{-3}$ and a more average case with a dissimilarity of $2.0\times 10^{-3}$. Even in the worst case example, we see that the level of similarity between them is sufficient for the GAN to learn a realistic speckle pattern. 

We perform image reconstruction on the unlabelled test set (440 images w/o known bright sources). The mean dissimilarity and its standard deviation is $(2.1 \pm 1.3)\times10^{-3}$ (so the worst case example shown in Fig. \ref{fig:gan_compare} is an outlier). Convinced that the GAN can generate adequately realistic imitations of the speckle pattern, we will now use it as a data generator for training a classifier.

\section{A CNN for supervised planet detection}

\subsection{Generating labelled synthetic datasets}
In order to create synthetic datasets suited to train image classifiers, we generated 10,000 synthetic images using the previously trained GAN model. Although these images are all examples of the negative class, we are able to produce for each of them a duplicate example in the positive class by injecting a planet signal. Because of this \emph{paired samples} approach, a classifier trained on these data is less likely to associate random features of the speckle pattern with the presence of a planet.


The planet signal is introduced by artificially injecting\footnote{We opted not to use a GAN for augmenting the positive examples (i) to fully control the SNR of the injected planets, for evaluation purposes and (ii) because the randomly positioned faint planet signal in positive examples would be easily masked by the most prevalent features of the images, i.e. those comprising the speckle pattern.} a simulated \emph{Point Spread Function (PSF)}\footnote{The PSF is the response of the telescope optics to incoming light, i.e. it defines the light distribution of a point-source, e.g. a planet, on the detector plane.} to the image. We used \emph{TinyTim} \cite{TinyTim}, an instrument-specific PSF simulator for NICMOS, to generate a normalised PSF. Although in real observations the shape of the PSF may differ slightly depending on the position of the light source, in this work we assume this difference to be negligible and thus use the same PSF regardless of the position of the injected planet. 

To generate a planet signal with a given SNR, we uniformly sample a pixel $P = (x, y)$ to be its center. We consider a $4\times4$ window centered\footnote{The window's top-left corner is $(x-1, y-1)$ and bottom-right is $(x+2, y+2)$.} at $P$ and we denote by $\sigma$ the standard deviation of the pixel intensity values in it, so that we can determine the injected signal's brightness. Under the definition of SNR, the total pixel intensity of the signal $S_p$ is computed by
\begin{equation*}
S_p = SNR \times \sigma \times (4\times4).
\end{equation*}
The intensity of the PSF to be injected in the image is thus determined by $S_p$ and affects the pixels within a $33\times33$ window centered at $P$. After the planet PSF injection, each synthetic image is normalised so that its pixel intensities lie in [0,1] using its minimum and maximum pixel values.


The process of injection depends only on the given SNR and the noise signal surrounding the sampled center. No assumptions were made regarding the planet’s separation from the star and its brightness. In order to avoid injecting planets brighter than the star, we impose an additional constraint: if the maximum pixel intensity of the PSF to be injected exceeds that of the original image (speckle pattern), then the signal center is sampled again.
In practice this means that certain areas in the center of the image never contain a planet for high SNRs.

Fixing the SNR allows training models at any desired level of `difficulty' and detecting planet signals over a wide range of brightness levels compared to their surrounding pixels. It also allows us to assess the limits of a classifier trained to detect objects at a given value of the SNR when deployed on a dataset of a different SNR.
To this end, we assembled $4$ datasets each obtained by fixing a different SNR level while producing the positive classes. The SNRs considered are 1.5, 1.25, 1 \& 0.75.

Before the artificial images are presented to the classifier, the four datasets are split coherently into training (80\%, 16000 images)
and test (20\%, 4000 images) sets. By coherent, we mean the split was applied once on the negative class (which is common among the datasets) and then extended to the positive classes in such a way that paired samples appear in the same set.

\subsection{Training the CNN}

%
%
%

\begin{table}[t]
\scriptsize
\centering
\begin{tabular}{c|c|c}
\hline
Layer Type   & Dimensionality, Configuration & Output Dimension \\ \hline \hline
Input        & 64$\times$64$\times$1       & (64,64)              \\
Conv-RELU-BN & 3$\times$3, f=8, s=1     & (m,64,64,8)      \\
MaxPool      & 2$\times$2, s=2       & (m,32,32,8)      \\
Conv-RELU-BN & 3$\times$3, f=16, s=1    & (m,32,32,16)     \\
MaxPool      & 2$\times$2, s=2       & (m,16,16,16)     \\
Conv-RELU-BN & 3$\times$3, f=32, s=1    & (m,16,16,32)     \\
MaxPool      & 2$\times$2, s=2       & (m,8,8,32)       \\
FC           & 256           & (m,256)          \\
Sigmoid      & 2             & (m,2)   \\\hline        
\end{tabular}
\small
\caption{Architecture of the CNN classifier trained on data with SNR values 1.5 \& 1.25. Naming convention follows that of Table \ref{tab:GAN_config}. For SNR values 1 \& 0.75 the only changes where that double convolutional layers were used instead of single ones and the FC layer consisted of 128 neurons.}
\label{tab:CNN_config_simple}
\vspace{-1cm}
\end{table}

\begin{table}[]
\resizebox{\textwidth}{!}{%
\centering
\begingroup
\setlength{\tabcolsep}{6.25pt} 
\begin{tabular}{cccccc}
 &  & \multicolumn{4}{c}{Test SNR} \\ \cline{3-6} 
 & \multicolumn{1}{c|}{Train SNR} & \multicolumn{1}{c|}{1.5} & \multicolumn{1}{c|}{1.25} & \multicolumn{1}{c|}{1} & \multicolumn{1}{c|}{0.75} \\ \cline{1-6} 
ACC & \multicolumn{1}{|c|}{\multirow{3}{*}{1.5}} & \multicolumn{1}{c}{0.972 $\pm$ 0.004} & \multicolumn{1}{c}{0.945 $\pm$ 0.006} & \multicolumn{1}{c}{0.879 $\pm$ 0.010} & \multicolumn{1}{c|}{0.752 $\pm$ 0.009} \\ 
TPR & \multicolumn{1}{|c|}{} & \multicolumn{1}{c}{0.943 $\pm$ 0.008} & \multicolumn{1}{c}{0.900 $\pm$ 0.012} & \multicolumn{1}{c}{0.759 $\pm$ 0.020} & \multicolumn{1}{c|}{0.503 $\pm$ 0.018} \\ 
TNR & \multicolumn{1}{|c|}{} & \multicolumn{4}{c|}{1.000 $\pm$ 0.000} \\ \cline{1-6} 
ACC & \multicolumn{1}{|c|}{\multirow{3}{*}{1.25}} & \multicolumn{1}{c}{0.973 $\pm$ 0.004} & \multicolumn{1}{c}{0.955 $\pm$ 0.003} & \multicolumn{1}{c}{0.900 $\pm$ 0.002} & \multicolumn{1}{c|}{0.786 $\pm$ 0.010} \\ 
TPR & \multicolumn{1}{|c|}{} & \multicolumn{1}{c}{0.949 $\pm$ 0.007} & \multicolumn{1}{c}{0.911 $\pm$ 0.005} & \multicolumn{1}{c}{0.803 $\pm$ 0.004} & \multicolumn{1}{c|}{0.575 $\pm$ 0.021} \\ 
TNR & \multicolumn{1}{|c|}{} & \multicolumn{4}{c|}{0.998 $\pm$ 0.003} \\ \cline{1-6} 
ACC & \multicolumn{1}{|c|}{\multirow{3}{*}{1}} & \multicolumn{1}{c}{0.973 $\pm$ 0.003} & \multicolumn{1}{c}{0.963 $\pm$ 0.003} & \multicolumn{1}{c}{0.938 $\pm$ 0.004} & \multicolumn{1}{c|}{0.862 $\pm$ 0.006} \\ 
TPR & \multicolumn{1}{|c|}{} & \multicolumn{1}{c}{0.955 $\pm$ 0.011} & \multicolumn{1}{c}{0.942 $\pm$ 0.012} & \multicolumn{1}{c}{0.897 $\pm$ 0.015} & \multicolumn{1}{c|}{0.752 $\pm$ 0.024} \\ 
TNR & \multicolumn{1}{|c|}{} & \multicolumn{4}{c|}{0.995 $\pm$ 0.004} \\ \cline{1-6} 
ACC & \multicolumn{1}{|c|}{\multirow{3}{*}{0.75}} & \multicolumn{1}{c}{0.967 $\pm$ 0.008} & \multicolumn{1}{c}{0.957 $\pm$ 0.009} & \multicolumn{1}{c}{0.939 $\pm$ 0.011} & \multicolumn{1}{c|}{0.896 $\pm$ 0.010} \\ 
TPR & \multicolumn{1}{|c|}{} & \multicolumn{1}{c}{0.947 $\pm$ 0.003} & \multicolumn{1}{c}{0.927 $\pm$ 0.004} & \multicolumn{1}{c}{0.890 $\pm$ 0.010} & \multicolumn{1}{c|}{0.804 $\pm$ 0.022} \\ 
TNR & \multicolumn{1}{|c|}{} & \multicolumn{4}{c|}{0.988 $\pm$ 0.019} \\ \cline{1-6} 
\end{tabular}
\endgroup
}
\caption{Test Accuracy (ACC), True Positive Rate (TPR) \& True Negative Rate (TNR) of the CNN classifier per training \& test SNR combination. We report mean \& standard deviation across 5 runs. The SNR applies to Positive examples only, so Negatives are the same for all SNR values on a given run. We see that the lower the SNR on which the CNN is trained, the better its predictive performance across the entire SNR spectrum.} 
\label{tab:CNN_result}
\vspace{-0.75cm}
\end{table}

The four synthetic datasets were used to train and to evaluate various CNN classifiers. The architectures of the CNNs used vary slightly on different SNR levels, as shown in Table \ref{tab:CNN_config_simple}. All models were trained using the \emph{ADAM} optimiser with default \emph{Keras} values, a learning rate of $10^{-4}$ and a batch size of 16. Dropout was applied to the final fully connected layers with a rate of 0.35. The maximum number of epochs was 100, but early stopping was applied if the validation loss was not decreasing for 20 epochs. The hyperparameter optimisation was based on minimising the validation loss and was minimal due to the already good performance of the final product. A principled hyperparameter exploration is left for future work.

For each of the 4 different levels of SNR examined (1.5, 1.25, 1 \& 0.75), we train a model on the 75\% of the corresponding training set (12,000 images) and we use the remaining 25\% as a validation set (4,000 images) to monitor the training performance. In the same way as before, this last split is performed coherently, in order to respect the samples' pairing. The training procedure was repeated 5 times for each dataset, applying different train/validation splits and thus producing 5 different models for each SNR considered.

Table \ref{tab:CNN_result} shows the performance of the CNNs trained on the same SNR when evaluated on the 4 different test sets (note that the coherency of the splits guarantees that the speckle patterns appearing in the test sets are always unseen data). We observe that the predictive performance is high and the models are robust to deployment on a different SNR than the one they were trained on -- especially if the training SNR is lower than the test SNR. This is reasonable, as CNNs trained on lower SNRs (i.e. trained to recognise fainter signals) are expected to perform well at classifying datapoints of higher SNRs (i.e. brighter signals). In light of this, any subsequent results presented are obtained by the CNN trained on SNR\,=\,0.75.

So far, we have assessed the classifiers' capability to detect planets on synthetic data generated from the same distribution as their training set. The fact that models trained at a given SNR can perform well even when the test set SNR is lower is encouraging; it means that the models we train might be able to detect faint planet signals. 

\section{Locating planets \& assessing sensitivity}
\subsection{Locating planets}
\vspace{-1cm}
\begin{figure}
    \centering
    \includegraphics[width = 0.8\linewidth]{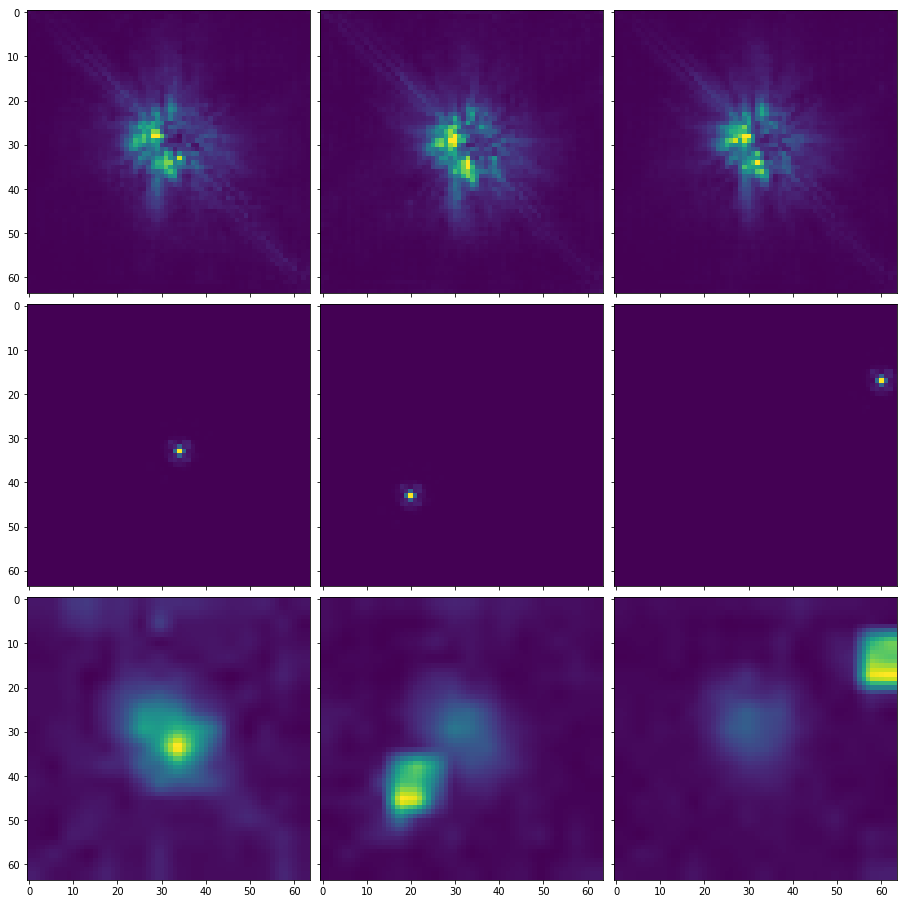}
    \caption{Locating planets on 3 exemplar synthetic datapoints [Left to Right]. [Top] Input synthetic image containing a planet; [Middle] Actual planet location in the image (contrast enhanced for visualisation); [Bottom] CNN Class Activation Map on the image.}
    \label{fig:TP_examples}
\end{figure}

Knowing whether a given image does, or does not, contain a planet does not provide much information on \emph{where} the candidate planet may be located. To answer this question, we visualised the \emph{Class Activation Maps (CAMs)} \cite{zhou2016learning} of the CNN to investigate which regions of the image contribute the most to its predictions. The CAM is obtained by taking the weighted average (w.r.t. their weight to the next layer) of the features in the final convolutional feature map.

As an initial test of our model's ability to locate planets, we applied this technique on its predictions on the synthetic test data (4,000 labelled images not seen by the CNN). Fig. \ref{fig:TP_examples} shows some examples of True Positives. For each example, we compare the actual location of the planet to the CAM of the CNN, highlighting the regions most inductive to its decision. We see that when the CNN classifies an image as one that contains a planet, it does so because it `spots' the planet. Thus, the CNN trained to classify images as potentially containing planets can also be used to locate potential planets within the images.

\begin{figure}
    \centering
    \includegraphics[width=\linewidth]{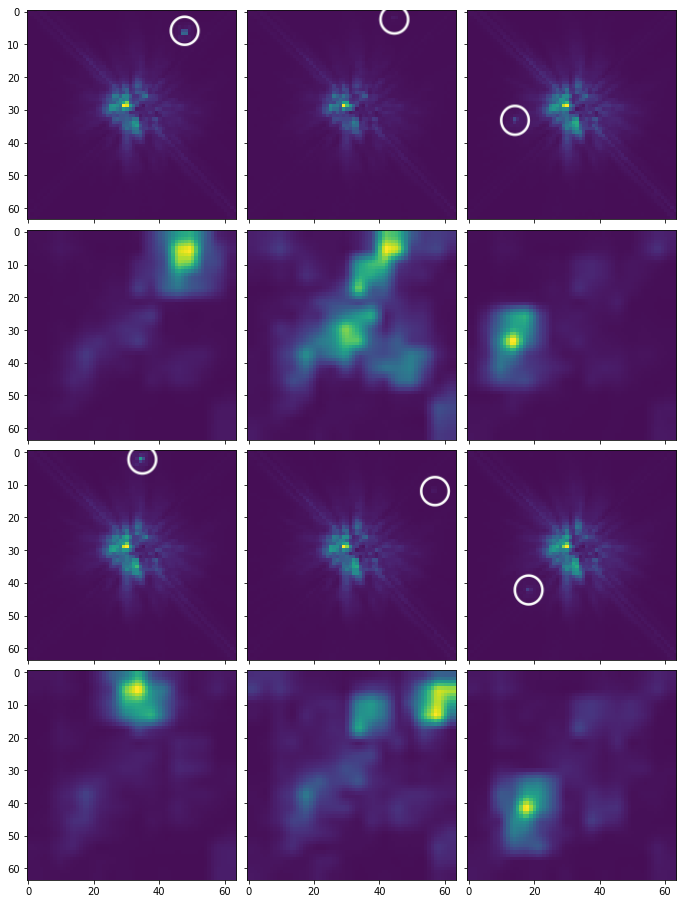}
    \caption{Original images of CQ-Tau [Left], DQ-Tau [Middle] \& HD10578 [Right], each in 2 orientations [1st Row \& 3rd Row]. The images contain confirmed bright sources (marked with a circle). [2nd Row \& 4th Row]: CNN activation heatmaps of model trained at SNR = 0.75; We see that the model's activation peaks on the bright sources.}
    \label{fig:confirmed}
\end{figure}

So far the network has performed well on synthetic data generated by the GAN. However, we have yet to answer if it can perform well on actual NICMOS data. Therefore, the next step is to perform the classification on held out images of targets with confirmed bright source detections. Tested on all positive held out real images (36 datapoints of 3 targets: \emph{CQ-Tau}, \emph{DQ-Tau} \& \emph{HD10578}) the model can detect the bright source in each image. In Fig. \ref{fig:confirmed}, we can see 2 example images with successful detections for each of the 3 targets. The positions of these bright sources (in circle) were independently detected by the ALICE program\footnote{\url{https://archive.stsci.edu/prepds/alice/ }}~\cite{Hagan2018}. 
 
\subsection{Assessing sensitivity \& specificity}

\begin{figure}[h]
\centering
\includegraphics[width=0.8\linewidth]{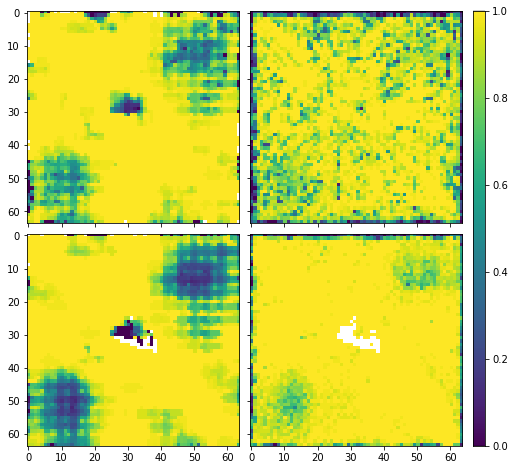}\\
\caption{Precision [Left] \& Recall [Right] map of planet localization per input image pixel. We show average results across 5 runs of the CNN trained on SNR=0.75 when tested on synthetic data of SNR=0.75 [Top] and SNR=1.5 [Bottom]. We consider it a True Positive when the planet's PSF center is one of the top 25\% most activating pixels of the CAM. To assign False Positives, we consider the brightest pixel of the CAM (corresponding to 16 pixels in the image) as the predicted planet center.}
\label{fig:precision_recall_map}
\end{figure}

Finally, in Fig. \ref{fig:precision_recall_map}, we visualise the precision and recall capability of the model to locate planets on each pixel of the 64$\times$64 images. The precision map shows how likely a planet is to really be centered at a pixel, given the model `detects' it. The recall map shows how likely a planet is to be detected by the model if it really is centered on that pixel.

For the purposes of this visualisation, we produced an extended synthetic test set for each SNR value, in which, for each of the 64$\times$64 pixels, up to 5 test instances (subject to dataset generation restrictions) had a planet PSF centered in that pixel. In Table \ref{tab:CNN_result}, we were only assessing image classification performance. So, if a model classified a positive example as one that contains a planet, this would count as a True Positive (TP), even if the features most contributing to the classification did not include the center of the planet PSF. Now, for the purposes of assessing planet localisation within the image, such an example will count as a False Positive (FP) for the pixel(s) falsely identified as the center of a planet PSF and a False Negative (FN) for the pixel being the actual center of the undetected planet PSF. We consider it a success (TP) when one of the top 25\% pixels activating the model was the center of the injected planet PSF. To assign FPs, we consider the brightest pixel of the CAM (corresponding to 16 pixels in the image) as the predicted planet center. We then calculate per pixel,
\begin{equation*}
Precision = \frac{TP}{TP+FP} \qquad  \text{and} \qquad Recall = \frac{TP}{TP+FN}.
\end{equation*}
Both precision and recall are very high in most areas of the image, except for the edges and the center. The edge effects are a result of the convolutions and can easily be avoided by considering larger frames. The blind spot in the center results from the speckle pattern itself and the fact that for high SNR values, in certain central areas, no datapoint contained a planet (see dataset generation; these areas are the pixels in white color on the maps). In real observations the 8$\times$8 pixel center of the image is occulted by the coronagraph, so in the innermost region of the speckle pattern it would be impossible to detect any companions.

We also see that planets lying on the main components of the speckle pattern (main diagonal \& `cross' centered on the image) are easier to detect and conversely, detections in these areas are more likely to be true planets. This effect appears to be an artifact of the dataset generation: as these are typically the brightest areas on an image, if a planet is to be placed there, to maintain a fixed SNR, it needs to be very bright. When the planet PSF is added to the background image, an already bright area becomes brighter and this is a feature the model picks up as a salient one for predicting the presence of a planet. Future extensions of this work will address this by defining the SNR in terms of planet/star contrast.

\vspace{-0.25cm}
\section{Conclusions \& Future Work}
\vspace{-0.2cm}
We have demonstrated that deep learning offers a promising direction for taking the field of direct imaging of exoplanets to the next step. By using a generative model of the data and injecting bright sources to generate a labelled synthetic dataset, we bypass the obstacles raised by directly using real data to train a classifier, namely the absence of positive examples and the possible presence of undetected planets in the data.

The supervised CNN classifier trained on the synthetic data can achieve high predictive performance and robustness to not only various levels of SNR, but also to SNR discrepancies between training and testing. Moreover, the model is capable of successfully locating astrophysical point sources within the images with the use of class activation maps. To demonstrate the limits of the model, we provided maps of the precision and recall of the model per pixel of the input image. When evaluated on actual data from NICMOS, the model can reproduce confirmed bright source detections in the data.

Our immediate next step is to apply this methodology to reconfirm the detection of known exoplanetary systems on images from a different filter of \emph{NICMOS}, \emph{FW160}. This will establish its ability to detect exoplanets and open the way for direct application of the method to unlabelled real data to identify new potential planet candidates. The most promising candidates can then be selected based on (i) the calibrated probability estimates of the CNN for each image to contain a planet and then (ii) weighing the heatmaps by the precision maps to pinpoint the most promising location within the images for the planets to be.

There is significant room for further improvement by tuning the architectures and the training hyperparameters of both the GAN used to generate the data and the CNN trained to detect planets. To make the synthetic dataset more realistic, future work will include factoring into the SNR the planet/star contrast, refining the PSF of planets injected at the edges of the image and injecting more than one planet per image at different values of SNR. To improve the planet-detectors themselves, the next steps include leveraging --when available-- multiple observations of the same target and combining planet-detectors to obtain an ensemble of predictors covering a wide range of SNR. The models presented here were trained on correctly classifying images as containing injected planets or not. As we saw, this decision was largely based on locating a planet or not, but ultimately, planet localisation was a byproduct of training. Future versions of our system will be trained directly on planet localisation within the images.

Finally, we will move beyond NICMOS and explore direct imaging datasets from different ground and space-based instruments, with a larger number of datapoints, more positive class examples and more observations per target. This will allow us to compare the `sensitivity' afforded by each instrument, at various levels of SNR and across different regions of an image, to use transfer learning methods and to combine observations from different instruments.


\vspace{-0.2cm}
\section*{Acknowledgments}
\vspace{-0.2cm}
 This project has received funding from the European Research Council (ERC) under the EU Horizon 2020 research \& innovation programme (grant No 758892, ExoAI) and under the EU Seventh Framework Programme (FP7/2007-2013)/ ERC grant No 617119 (ExoLights). Furthermore, we acknowledge funding by the Science \& Technology Funding Council (STFC) grants: ST/K502406/1, ST/P000282/1, ST/P002153/1 and ST/S002634/1. We are grateful for the support of the NVIDIA Corporation through the NVIDIA GPU Grant program.

\bibliographystyle{splncs04}

 \end{document}